\documentclass[prl,twocolumn,showpacs,superscriptaddress]{revtex4-1}
\usepackage{amsmath,amssymb,amsfonts}
\usepackage{graphicx}
\usepackage{setspace}
\usepackage{color}
\usepackage{txfonts}

\definecolor{light-gray}{gray}{0.5}

\newcommand{\citeData}[1]{\cite{dataPackage}#1}

\usepackage[unicode,breaklinks]{hyperref}
\hypersetup{
    unicode=true,
    a4paper=true,
    plainpages=false,
    pdftitle={Quantum reflection of bright solitary matter-waves from a narrow attractive potential},
    pdfauthor={},
    pdfsubject={Quantum reflection of bright solitary matter-waves from a narrow attractive potential},
    colorlinks=true,
    linkcolor=blue,
    citecolor=blue,
    filecolor=black,
    urlcolor=blue
}
\newcommand{\eqnrefp}[1]{{[Eq.~(\ref{#1})]}}

\newcommand{\eqnreft}[1]{{Eq.~(\ref{#1})}}

\newcommand{\figreft}[2]{Fig.~\ref{#1}#2}

\newcommand{\figrefp}[2]{[Fig.~\ref{#1}#2]}

\begin{document}

\title{Quantum reflection of bright solitary matter-waves from a narrow attractive potential}

\author{A.~L.~Marchant}
\affiliation{Joint Quantum Centre (JQC) Durham-Newcastle, Department of Physics, Durham University, Durham DH1 3LE, United Kingdom}
\author{T.~P.~Billam}
\affiliation{Joint Quantum Centre (JQC) Durham-Newcastle, Department of Physics, Durham University, Durham DH1 3LE, United Kingdom}
\author{M.~M.~H.~Yu}
\affiliation{Joint Quantum Centre (JQC) Durham-Newcastle, Department of Physics, Durham University, Durham DH1 3LE, United Kingdom}
\author{A.~Rakonjac}
\affiliation{Joint Quantum Centre (JQC) Durham-Newcastle, Department of Physics, Durham University, Durham DH1 3LE, United Kingdom}
\author{J.~L.~Helm}
\affiliation{Joint Quantum Centre (JQC) Durham-Newcastle, Department of Physics, Durham University, Durham DH1 3LE, United Kingdom}
\author{J.~Polo}
\affiliation{Department de F\'isica, Universitat Aut\`onoma de Barcelona, E-08193 Bellaterra, Spain}
\author{C.~Weiss}
\affiliation{Joint Quantum Centre (JQC) Durham-Newcastle, Department of Physics, Durham University, Durham DH1 3LE, United Kingdom}
\author{S.~A.~Gardiner}
\affiliation{Joint Quantum Centre (JQC) Durham-Newcastle, Department of Physics, Durham University, Durham DH1 3LE, United Kingdom}
\author{S.~L.~Cornish}
\email{s.l.cornish@durham.ac.uk}
\affiliation{Joint Quantum Centre (JQC) Durham-Newcastle, Department of Physics, Durham University, Durham DH1 3LE, United Kingdom}

\date{\today}

\pacs{
03.75.Lm,     	
03.75.-b,			
03.75.Kk			
}

\begin{abstract}

We report the observation of quantum reflection from a narrow, attractive,
potential using bright solitary matter-waves formed from a $^{85}$Rb
Bose--Einstein condensate. We create narrow potentials using a tightly focused,
red-detuned laser beam, and observe reflection of up to 25\% of the atoms,
along with the trapping of atoms at the position of the beam. We show that the
observed reflected fraction is much larger than theoretical predictions for a
narrow Gaussian potential well; a more detailed model of bright soliton propagation,
accounting for the generic presence of small subsidiary intensity maxima in the
red-detuned beam, suggests that these small intensity maxima are the cause of
this enhanced reflection.    

\end{abstract}

\maketitle

%
Solitons are non-dispersive and self-localised wave solutions that arise when
nonlinear interactions are sufficient to overcome the wavepacket
dispersion. Since the first observations in shallow water~\cite{Russell1845},
extensive studies of such solitary wave solutions have been carried out in a
diverse range of fields, including nonlinear optics and optical
fibers~\cite{Mollenauer1980,Kartashov2011,Kivshar2003}, plasma
physics~\cite{Stasiewicz2003} and magnetism~\cite{Togawa2013}. In the context
of quantum gases, quasi-one-dimensional (1D) Bose--Einstein condensates (BECs)
may be well described by the homogeneous 1D Gross-Pitaevskii equation (GPE), a
nonlinear Schr\"{o}dinger equation that manifests exact bright soliton solutions
for attractive interatomic interactions, taking the form of
localised density maxima~\cite{Kevrekidis2008}. Experimentally, a quasi-1D limit is approached by confining the condensate in a
highly elongated trap with tight radial confinement. While such traps
typically also feature weak axial confinement, precluding mathematically
exact soliton solutions, the resulting solitary wave solutions retain many
characteristics of the ideal soliton~\cite{Carr2002,Billam2012,Billambook}.
Previous experimental work has realized both single and multiple bright
solitary matter-waves using $^{7}$Li
atoms~\cite{Khaykovich2002,Strecker2002,Medley2014} and $^{85}$Rb
atoms~\cite{Cornish2006,Marchant2013,McDonald2014}, 
stimulating intense theoretical interest (see~\cite{Billambook} and
references therein).

\begin{figure}
\centering
\includegraphics[width=\columnwidth]{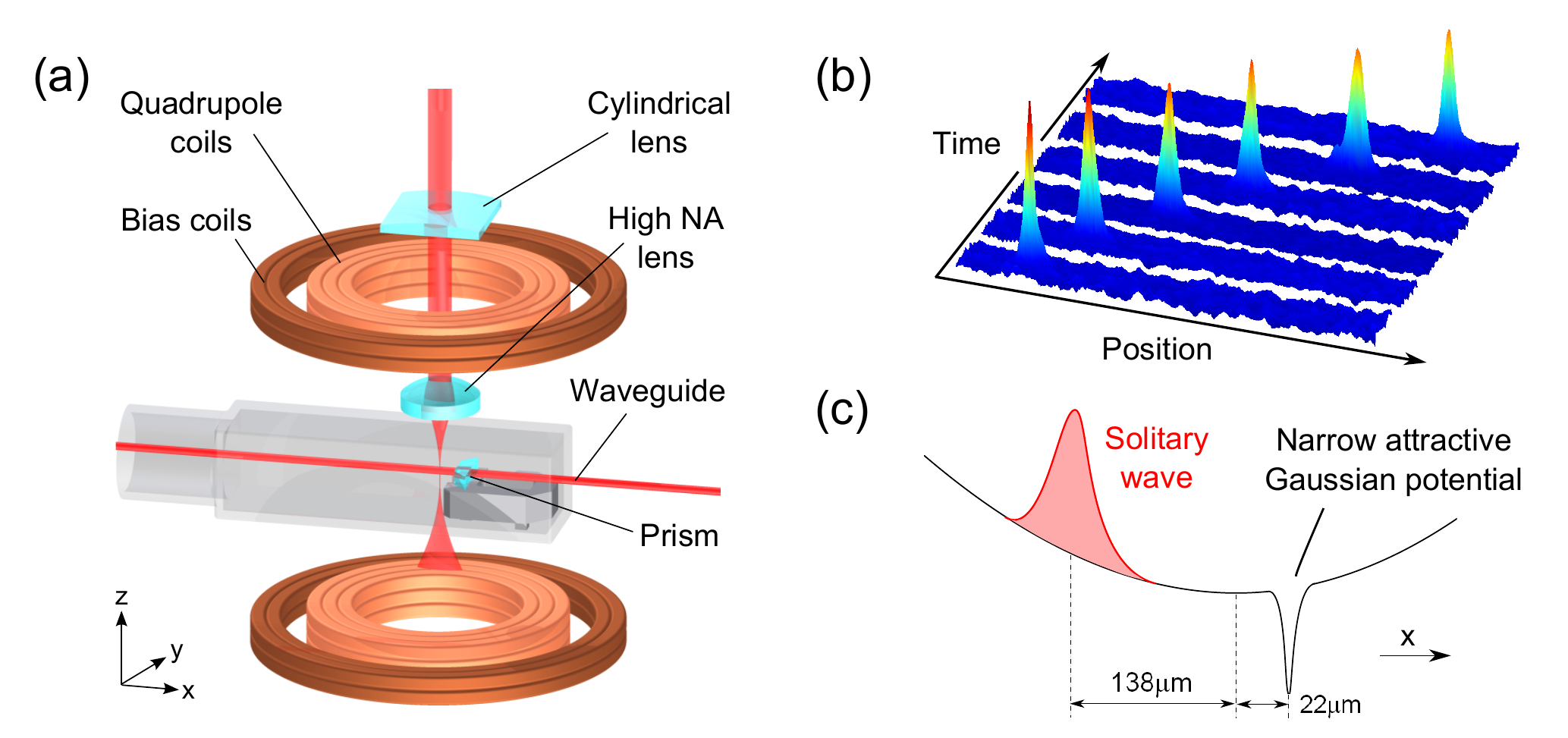}
\caption{(Color online) (a) Experimental setup. Atoms are cooled in a
crossed optical dipole trap (not shown), and then transferred into an
optical waveguide. Additional axial confinement is provided by magnetic
quadrupole and bias fields. The narrow attractive potential is formed
using a high numerical aperture (NA) lens to
produce a light sheet, tightly focussed in the $x$ direction. (b) Absorption
images of solitary wave propagation in the optical waveguide. (c) Schematic
showing the position of the narrow attractive potential, relative to the trap
center and the initial position of the solitary wave.
\label{fig:barrier_setup}}
\end{figure}

Scattering of bright solitary matter waves from narrow \textit{repulsive} 
potential barriers has been extensively studied theoretically in the context of 
matter-wave
interferometry~\cite{Helm2012,Martin2012,Cuevas2013,Polo2013,Helm2014,Helm2015}.
The nature of the scattering depends crucially on the center-of-mass kinetic
energy of the solitary wave relative to the modulus of its ground state energy.  For
high kinetic energies the barrier can act as a beam splitter; the two resulting
solitary waves can then be recombined to form an
interferometer~\cite{Helm2012}, with the outcome of the recombination depending
strongly on their relative phase~\cite{Parker2008} (as recently experimentally
demonstrated~\cite{Nguyen2014}). For low kinetic energies, the
scattering of the matter-wave soliton from the barrier can produce quantum
superposition states~\cite{Weiss2009,StreltsovEtAl2009b,Gertjerenken2012}.
Previous theoretical studies have also addressed the scattering of bright
solitary waves from narrow \textit{attractive} potential wells, where the
possibility exists for the bright solitary wave to undergo quantum reflection.
Depending on the parameter regime, significant quantum reflection has been
predicted for low energy solitons~\cite{Lee2006}, and significant resonant
trapping when the attractive potential is capable of supporting a number of
bound states~\cite{Ernst2010}.

In this Letter, we report the observation of splitting and quantum reflection 
of a bright solitary matter-wave from a narrow, attractive potential formed
from a tightly focused, red-detuned laser beam.  We investigate how the
fraction of atoms reflected varies with the depth of the attractive potential,
and observe atoms trapped at the position of the well.  Surprisingly, we
measure much greater reflected fractions than can be explained by theoretical
predictions for a Gaussian potential well. We address this discrepancy via
extensive theoretical modeling using the GPE, providing strong evidence that
the presence of small subsidiary diffraction maxima in the red-detuned beam,
creating a multiple-well structure, is the main source of the enhanced
reflection. While small subsidiary diffraction maxima are generically
expected in tightly-focused beams, our experiment is unusual in that they cause
qualitative changes in behavior.  Our results suggest that carefully engineered
attractive multi-well potentials may make robust beamsplitters for solitary
wave interferometry.


In this work we create stable $^{85}$Rb condensates using the method described in \cite{Marchant2012}. We use the broad Feshbach resonance at $155\,$G between
atoms in the $F=2, m_F=-2$ state~\cite{Claussen2003} to tune the scattering length to positive values, avoiding the large negative background scattering length and the associated collapse instability \cite{Ruprecht1995, Roberts2000, Donley2001}. Our setup uses a levitated
crossed optical dipole trap~\cite{Jenkin2011} 
\footnote{The magnetic gradient required to
levitate $^{85}$Rb atoms in the $F=2, m_F=-2$ state is
$B_z'=22.5\,\mathrm{G\,cm}^{-1}$, however, it is found to be beneficial during
evaporation to reduce the gradient slightly, to $ B_z'\approx
21.5\,\mathrm{G\,cm}^{-1}$ to allow atoms to be removed from the trap more
easily.}, 
providing independent control of the trapping frequencies
(dominated by the optical confinement) and the magnetic bias which is used to tune the scattering length (with a sensitivity $\sim40\,a_0\,\mathrm{G}^{-1}$ around the zero crossing of the Feshbach resonance).  We produce nearly pure condensates of up to
$4\times10^4$ atoms at a scattering length of $a_s \approx 200~a_0$ in an
almost spherical trapping geometry,
$\omega_{x,y,z}=2\pi\times\{30(1),30(1),42(2)\}$\,Hz. The
condensate number is reduced to $\sim6000$ atoms by further evaporation to facilitate solitary wave production.

We form a bright solitary wave by loading the BEC into an optical
waveguide created by an additional dipole trapping beam~\cite{Marchant2013} as
shown in Fig.~\ref{fig:barrier_setup}(a). We first ramp the
scattering length to a small positive value, $\sim 5~a_0$, over 50\,ms. We then
simultaneously ramp the crossed dipole beams off and the waveguide beam on
in 250\,ms. At the same time, we increase the magnetic field gradient to
exactly levitate the atoms, and ramp the bias field to give a scattering
length of $a_s=-7~a_0$. This value of $a_s$ minimises the
dispersion of the condensate as it travels along the waveguide, see
Fig.~\ref{fig:barrier_setup}(b). After the loading is completed the BEC is
confined radially in the waveguide beam but is free to propagate along the
axial direction. The combination of the magnetic field gradient, $B'_z$, and
the magnetic bias field, $B_z$, produces a weak harmonic potential in the axial
direction, given by $\omega_x=1/2\sqrt{\mu B_z'^2/mB_z}$, where $\mu$ is the
magnetic moment of the atoms, and $m$ their mass~\cite{Lin2009}. This magnetic potential
dominates the weak ($<0.1$~Hz) optical potential of the waveguide in the axial
direction yielding overall trapping frequencies of
$\omega_{x,y,z}=2\pi\times\{1.15(5),18.2(5),18.2(5)\}\mathrm{\,Hz}$. By carefully positioning the magnetic potential minimum with respect to the crossed dipole trap we are able to control the motion of the atoms in the waveguide.

\begin{figure}
	\centering
	\includegraphics[width=0.9\columnwidth]{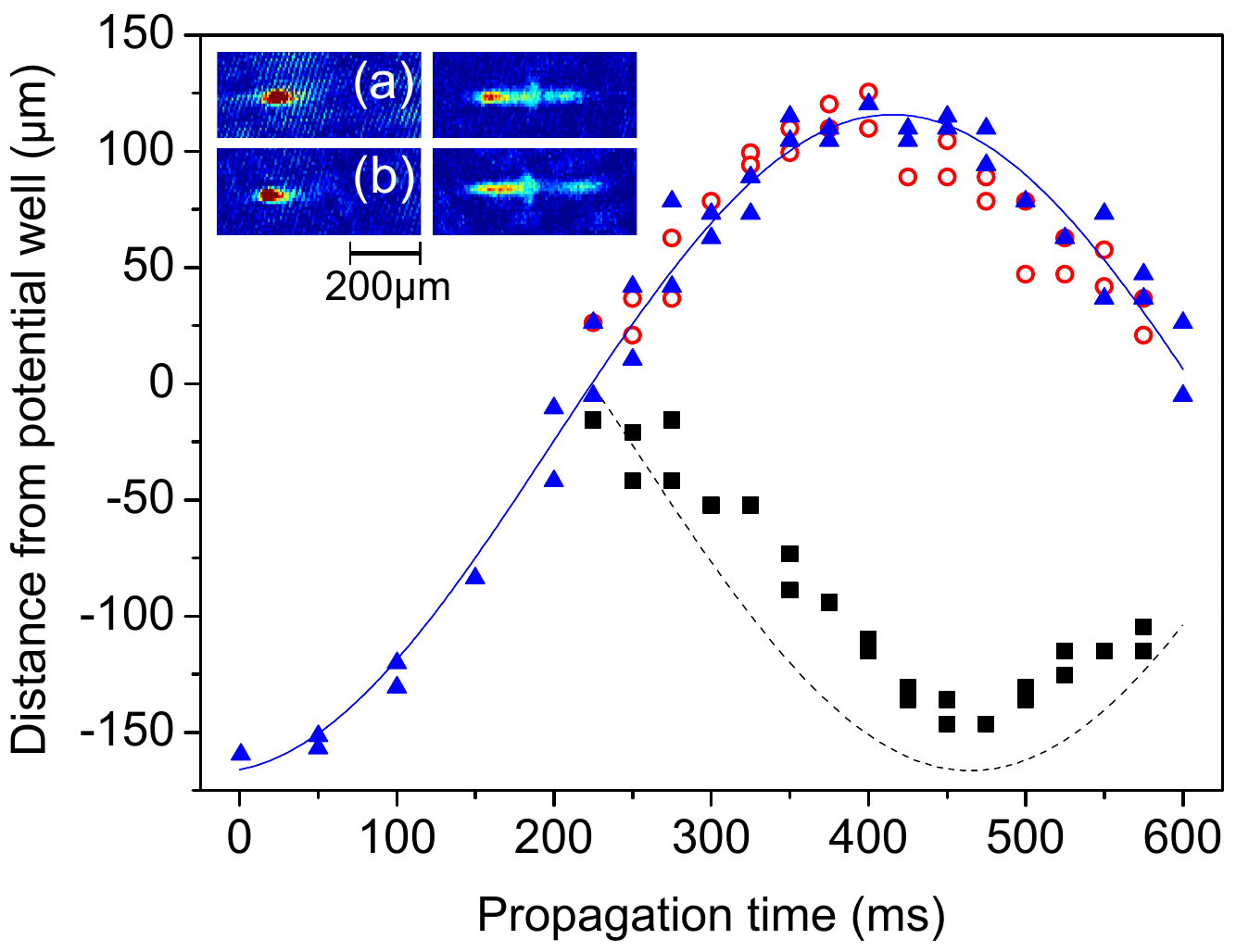}
	\caption{(Color online) Solitary wave position as a function of time. In the absence of the well (blue triangles) the atoms oscillate in the waveguide. With the well present the solitary wave splits, with atoms being both transmitted (red circles) and reflected (black squares). Lines indicate classical trajectories for free propagation (solid) and elastic reflection (dashed). Inset: False colour images taken at (a) 375 ms and (b) 475 ms with the well present (right) and absent (left).}
	\label{fig:Position_vs_time}
\end{figure}

We produce a narrow attractive potential well using $\lambda=852$\,nm light, focussed to form a light sheet and
determine the beam waists of $w_{x}
= 1.9(2)\,\mu$\,m and $w_{y} =570(40)\,\mu$\,m via parametric heating of atoms
trapped at the focus of the beam.  The potential well is precisely aligned with
respect to the waveguide in the vertical direction by mounting the final
lens in a threaded mount with a pitch of $1.4\,\mu\mathrm{m\,deg}^{-1}$. We position 
 the potential well $\sim 22\,\mu$m from the minimum of the axial
waveguide potential and release the solitary waves from the crossed dipole trap
situated $\sim 160\,\mu$m away from the well [as shown in Fig.~\ref{fig:barrier_setup}(c)],
giving an incident velocity of $\sim 1$\,mm\,s$^{-1}$. At full power we obtain a maximum well depth of $1\,\mu\,\mathrm{K}\times k_{\rm{B}}$.


In our initial experiment we set the potential well depth to its maximum value, release a solitary wave into the
waveguide, and track its position by imaging multiple instances of the same
experimental sequence at different times after release. Once the solitary wave
reaches the well, we observe a splitting of the wavepacket and identify three
distinct resulting fragments: atoms transmitted, reflected, and confined at the
potential well. We are able to track the center-of-mass positions of both the
transmitted and reflected atomic clouds, as shown in
Fig.~\ref{fig:Position_vs_time}. The majority of atoms in the solitary wave are
transmitted (red circles), following the same trajectory as in the freely
propagating case (blue triangles), undergoing harmonic motion in the waveguide
(solid line). Around $5-10~\%$ of the atoms are confined close to the well. The remainder of the atoms ($\sim25~\%$) reflect from the narrow potential
well and propagate in the opposite direction to the transmitted component. The
turning point of the reflected atoms occurs $\sim50$\,ms later than for the
transmitted atoms, due to the offset of the well position from the trap
center. This turning point is $\sim20\,\mu$m short of the release
position, suggesting some energy is lost during the splitting process. For comparison, the trajectory of an elastic collision is shown by the dashed line in Fig.~\ref{fig:Position_vs_time}. 

\begin{figure}
	\centering
	\includegraphics[width=0.9\columnwidth]{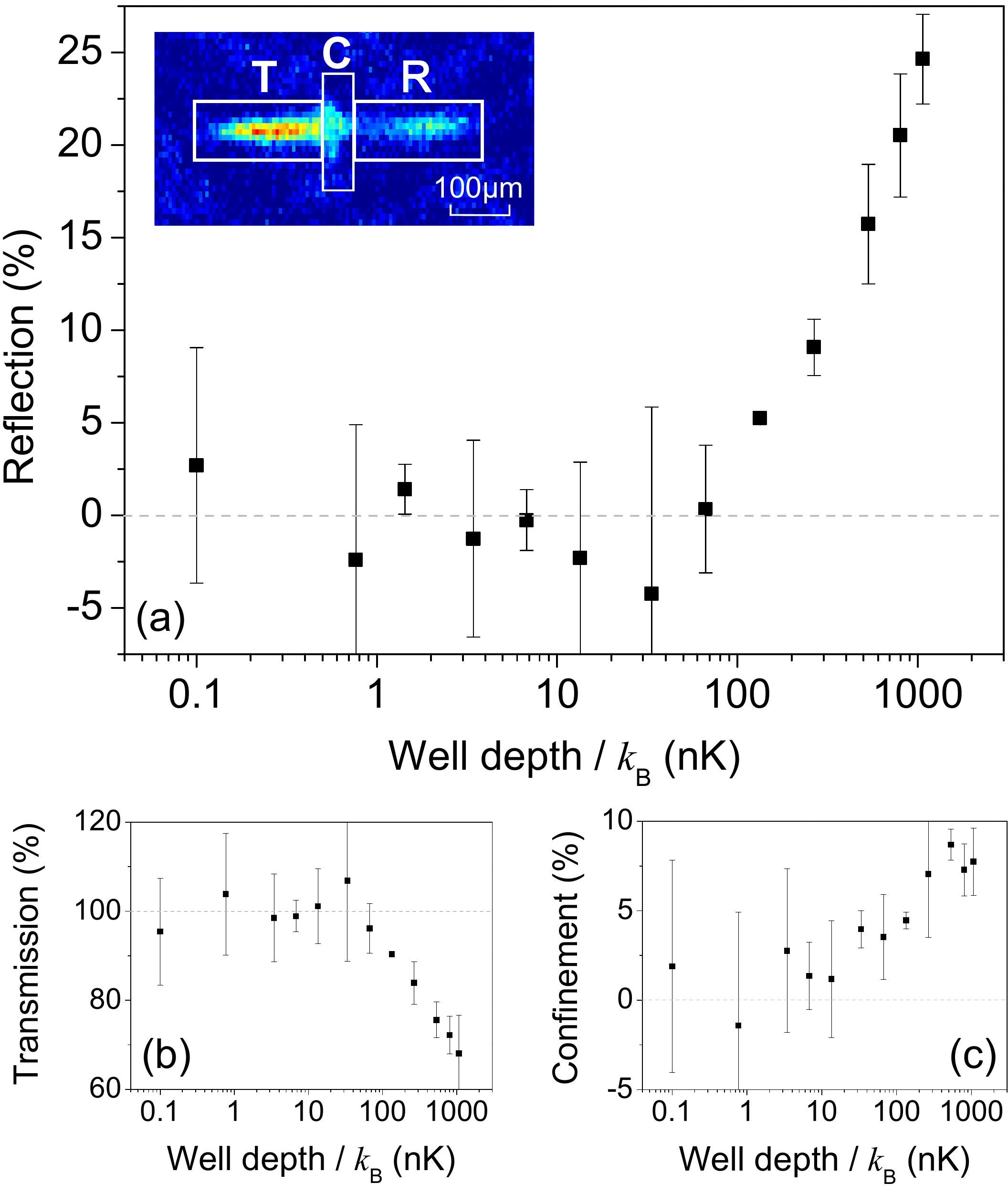}
	\caption{(Color online) The percentage (a) reflection (R), (b) transmission (T), and (c) confinement (C), of atoms as a function of well depth for an incident solitary wave with a velocity of 1\,mm\,s$^{-1}$. These percentages are determined using regions defined in the inset of (a). See text for details.}
	\label{fig:R_vs_power}
\end{figure}

To explore the effect of the potential well depth relative to the energy of the
incoming solitary wave we vary the power of the 852~nm beam, while keeping all
other parameters constant. The solitary wave is split and the resulting
fragments allowed to spatially separate before they are imaged, 475\,ms after
release. To calculate the reflection probability, we define three fixed regions
of the absorption images: transmitted (T), confined (C), and reflected (R), as
shown in the inset of Fig.~\ref{fig:R_vs_power}(a). Taking the sum of the pixel
values in each of these regions we define the reflection probability as
$\mathrm{R}/(\mathrm{R}+\mathrm{C}+\mathrm{T}) \times 100\%$ (values for the
transmitted and confined parts are calculated similarly) \footnote{Pixel values from shots with no atoms present are subtracted from the images to normalise the background noise level. This can lead to some reflection values below 0~\%, however, the error bars associated with this calculation means these values are still consistent with zero.}.  We find there is no
observable reflection from the narrow potential well for trap depths $<
100\,$nK. Above this threshold, the probability of reflection increases sharply
[see Fig.~\ref{fig:R_vs_power}\,(a)], and the number of atoms transmitted drops
correspondingly [Fig.~\ref{fig:R_vs_power}\,(b)]. For a trap depth of
$1\,\mu\mathrm{K}\times k_{\rm{B}}$, we observe a reflection of $\sim 25$\%. The
number of atoms confined at the position of the well also increases with
increasing well depth, as shown in Fig.~\ref{fig:R_vs_power}\,(c).

Intriguingly, the observed coefficient of reflection [Fig.\ref{fig:R_vs_power}(a)]
is too large to be explained in terms of quantum reflection from a Gaussian potential well, 
\begin{equation}
V_\mathrm{G}(x) =-V_0 \exp\left(-2x^2/\ell^2\right)\label{eqn:gauss}\,,
\end{equation}
where $V_0>0$ and $\ell = 1.9$ $\mu$m.  A simple approximate argument for this comes from
the analytic formula for the single-particle reflection coefficient for the
similar potential $V(x) = -V_0/\cosh^2(x/d)$ (choosing $d\approx \ell/1.6$)
\cite{LandauLifshitz2000}
\begin{equation}
R = \frac{\cos^2\left(\pi \sqrt{1/4+{2mV_0d^2}/{\hbar^2}}\right)}{\sinh^2(\pi kd) + \cos^2\left(\pi\sqrt{1/4+{2mV_0d^2}/{\hbar^2}}\right)}, \label{eqn:spformula}
\end{equation}
where $k$ is the wavevector of the incoming plane wave. Since $\cos^2(x) \le 1$ for all real arguments $x$, this approximation shows that
for $\ell=1.9$ $\mu$m a small incoming velocity (small $k$) is necessary to
observe any reflection, \textit{regardless of the
well depth} $V_0$. For velocities $\rm{v} \approx 1\mathrm{mm
}\,\mathrm{s}^{-1}$, as realized in the experiment, this approximation predicts negligible reflection. 

\begin{figure}[!t]
\centering
\includegraphics[width=\columnwidth]{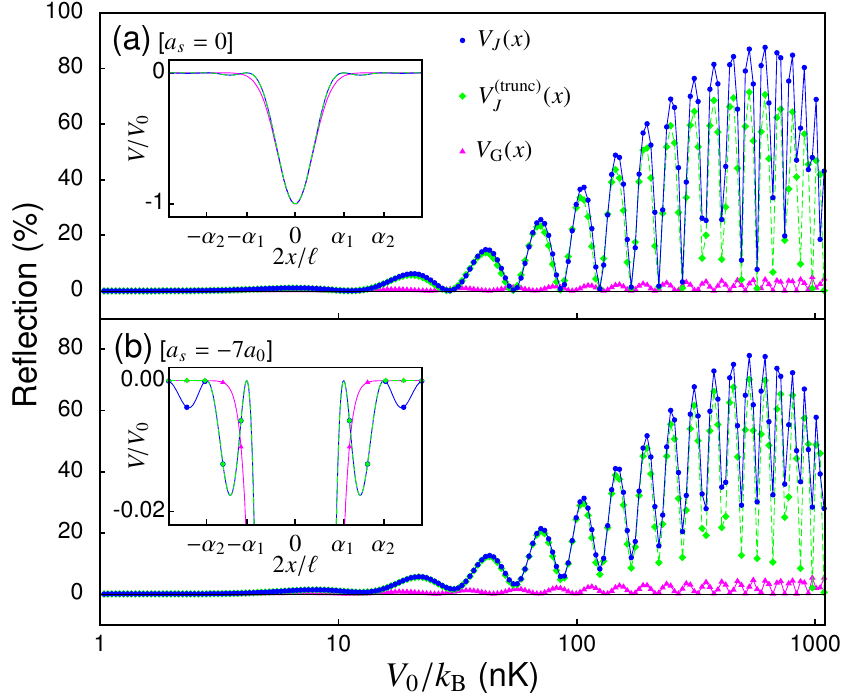}
\caption{(Color online) Reflection coefficients as a function of potential well depth for
non-interacting wavepackets (a) and bright solitary waves (b) in a 1D GPE model. Results are shown for a Gaussian potential
well with a single potential minimum [$V_\mathrm{G}(x)$], and for
diffraction pattern [$V_J(x)$] and truncated diffraction pattern
[$V_J^{(\mathrm{trunc})}(x)$] potentials with subsidiary minima (see insets). \label{fig:theory}}
\end{figure}

This lack of substantial reflection from the Gaussian potential
$V_\mathrm{G}(x)$ \eqnrefp{eqn:gauss} is confirmed by detailed numerical
simulations of a quasi-one-dimensional GPE 
\begin{equation}
i\hbar \frac{\partial \psi(x,t)}{\partial t} = \left[ \frac{-\hbar^2}{2m} \frac{\partial^2}{\partial x^2} + V(x) + U(x,t) + g_\mathrm{1D} |\psi(x,t)|^2 \right] \psi(x,t),
\end{equation}
where $U(x,t)$ represents the time-dependent background potential (see lower curves in \figreft{fig:theory}{}). We model the
latter as
\begin{equation}
U(x,t) = \frac{1}{2} m \left[ \omega_{x_1}(t)^2 (x-x_1)^2 + \omega_{x_2}(t)^2 (x-x_2)^2 \right],
\end{equation}
where $x_1 = -160$~$\mu$m ($x_2 = -22 $ $\mu$m) represents the location of the
minimum of the dipole (waveguide) potential in $x$, see Fig.~\ref{fig:barrier_setup}~(c). The trap frequencies for
these potentials are ramped linearly over the first $\tau=250$~ms:
$\omega_{x_1}(t) = \mathrm{max}\{2\pi\nu_1(\tau-t)/\tau,0\}$ and
$\omega_{x_2}(t) = \mathrm{min}\{2\pi\nu_2 t/\tau,2\pi\nu_2\}$, for
$\nu_1=30$~Hz and $\nu_2=1.15$~Hz. The (static) potential well $V(x)$ is
centered on $x=0$ in these coordinates, and the atoms move towards positive
$x$.  The nonlinearity $g_\mathrm{1D} = 4\pi N a_\mathrm{s} \hbar \nu_\perp$,
where we take $N=6000$ and $\nu_\perp = 18.2$~Hz. We work with $\psi(x,t)$
normalized to unity, and initialize the simulation with $\psi(x,t)$ in the
ground state of the system for potential $U(x,t=0)$.  In agreement with the
approximate formula [Eq.~(\ref{eqn:spformula})], these simulations confirm that
only very weak reflection ($\lesssim 4$\%) is expected from the Gaussian
potential $V_\mathrm{G}(x)$, both for non-interacting wavepackets
\figrefp{fig:theory}{(a), $a_\mathrm{s}=0$}, and for bright solitary waves
\figrefp{fig:theory}{(b), $a_\mathrm{s}=-7a_0$}. We have confirmed that the amount of reflection is not significantly changed by the use of a fully three-dimensional GPE model, the inclusion of noise in the initial wavepacket, or both of these extensions.

To qualitatively explain the surprisingly large observed reflections we consider the effects of subsidiary diffraction maxima
in the optical intensity, which occur generically
in focusing optical configurations \cite{BornAndWolf}. Since these are generally
much less intense than the primary maximum, they are
typically ignored when calculating optical potentials in BEC experiments.
However, in the context of our experiment, the \textit{narrow} nature of the
subsidiary maxima is potentially significant; at least when considered in isolation, they are able to produce larger reflection than the primary
maximum [see \eqnreft{eqn:spformula}]. Also, the presence of multiple potential
wells can itself enhance reflection; this is seen, for example, in Bragg
reflection of BECs from a multiple-well lattice \cite{Fabre2011a} (although in
our case the rapid variation in well depths precludes a similar analysis).

While the exact structure of the subsidiary diffraction maxima in the
red-detuned beam is not precisely known in our experiment, as a generic model
we consider the potential due to the intensity pattern of Fraunhofer
diffraction from an aperture \cite{BornAndWolf}
\begin{equation} 
V_J(x) = -V_0\left[ \frac{\ell}{x} J_1 \left(\frac{2x}{\ell}\right) \right]^2\,,
\end{equation}
and the same potential
truncated after the first subsidiary maxima; 
\begin{equation}
V_J^{(\mathrm{trunc})}(x) = \left\{ \begin{array}{cc} V_J(x) & |2x/\ell|<\alpha_2, \\ 0 & |2x/\ell| \ge \alpha_2, \end{array} \right.,
\end{equation} where $\alpha_2$ is the second positive zero of the Bessel
function $J_1(x)$.  As shown in \figreft{fig:theory}~{(inset)} these potentials
have a similar central minimum to $V_\mathrm{G}(x)$, but also one
[$V_J^{(\mathrm{trunc})}(x)$] or a decaying series [$V_J(x)$] of subsidiary
minima. The results of 1D GPE simulations for both noninteracting wavepackets
\figrefp{fig:theory}{(a), $a_\mathrm{s}=0$} and for bright solitary waves
\figrefp{fig:theory}{(b), $a_\mathrm{s}=-7a_0$} show that the reflection is
greatly enhanced for both of these potentials compared to $V_\mathrm{G}(x)$,
for the range of well depths used in the experiment. The presence of subsidiary
diffraction maxima in the beam producing the potential well thus provides a
plausible explanation for the non-zero reflection probabilities observed in the
experiment. The similarity of the results for $V_J(x)$ and
$V_J^\mathrm{(trunc)}(x)$ indicate that the oscillatory structure of the
reflection coefficient is primarily a transmission resonance effect,
attributable to the three-well potential composed of the main beam maximum and
the largest two subsidiary diffraction maxima. 

There are quantitative differences between the experimental data
[Fig.~\ref{fig:R_vs_power}(a)] and this generic model; in particular, the model
exhibits negligible ($<1$\%) confinement, and oscillatory structure. We have
excluded small shot-to-shot changes in the incoming soliton velocity due to
small ($\sim \pm 5~\mu$m) shifts in the alignment of the experimental
potentials as an explanation for the lack of oscillations in the experiment;
changing the initial displacement of the soliton by up to $\pm 5$ $\mu$m in the
model leads to a negligible change in reflection coefficient. We suspect that
these remaining differences may arise from effects not captured by our
one-dimensional Gross-Pitaevskii model, such as the exact structure of the
potential well (possibly including time-dependent fluctuations),
three-dimensional effects, and finite-temperature effects.

In summary, we have observed quantum reflection of a bright solitary matter-wave from a narrow, attractive potential, formed by a tightly focused laser beam. Reflection probabilities of up to $25\%$ are measured, with the remaining atoms either transmitted or trapped at the position of the potential well. Modeling of the system suggests that the exact spatial form of the potential well is crucial in determining the amount of reflection observed, with the presence of multiple optical diffraction maxima, rather than a single Gaussian maximum, playing an essential role. These results indicate that carefully engineered attractive multi-well potentials could be developed as robust beamsplitters for use in solitary wave interferometry. In future work we plan to replace the focused laser beam with a room-temperature super-polished glass prism (shown in Fig.~\ref{fig:barrier_setup}), allowing us to explore quantum reflection due to the attractive Casmir-Polder potential \cite{Cornish2009}.

The data presented in this paper are freely available to download~\citeData{}.

\acknowledgments
We thank Ifan Hughes and Matthew Jones for useful discussions. We acknowledge the UK Engineering and Physical Sciences Research Council (Grant No.\
EP/L010844/1, Grant No.\ EP/K030558/1) for funding. T.P.B. acknowledges financial
support from the John Templeton Foundation via the Durham Emergence Project
(http://www.dur.ac.uk/emergence).  J.P. acknowledges financial support from the FPI grant BES-2012-053447 and the mobility grant EEBB-I-14-08515, associated to the project FIS2014-57460-P, and from the Catalan government under the SGR 2014-1639 grant.



%

\end{document}